\documentclass[
    ,final            
  ]
  {aipproc}

\layoutstyle{6x9}

\usepackage{axodraw}
\usepackage{color}
\usepackage{graphicx}

\newcommand{\met}{\rlap{\,/}E_T}

\begin{document}

\title{Phenomenology of Universal Extra Dimensions}


\author{Kyoungchul Kong\footnote{current address: Theoretical Physics Department, Fermilab, Batavia, IL 60510, USA}}{
  address={Physics Department, University of Florida, Gainesville, FL 32611, USA}
}
\author{Konstantin T.~Matchev}{
  address={Physics Department, University of Florida, Gainesville, FL 32611, USA}
}

\begin{abstract}
In this proceeding, the phenomenology of Universal Extra Dimensions (UED), in which 
all the Standard Model fields propagate, is explored. We focus on models 
with one universal extra dimension, compactified on an $S_1/Z_2$ orbifold. 
We revisit calculations of Kaluza-Klein (KK) dark matter 
without an assumption of the KK mass degeneracy including all possible coannihilations.
We then contrast the experimental signatures of low energy supersymmetry 
and UED.
\end{abstract}

\maketitle


\section{Introduction}

Models of UED place all Standard Model particles in the bulk of 
one or more compactified extra dimensions. In the simplest and most popular
version, there is a single extra dimension of size $R$,
compactified on an $S_1/Z_2$ orbifold~\cite{Appelquist:2000nn}.
A peculiar feature of UED is the conservation of
Kaluza-Klein number at tree level, which is a simple consequence
of momentum conservation along the extra dimension.
However, bulk and brane radiative effects~\cite{Cheng:2002iz}
break KK number down to a discrete conserved quantity,
the so called KK parity, $(-1)^n$, where $n$ is the KK level.
KK parity ensures that the lightest KK partners 
(those at level one) are always pair-produced
in collider experiments, just like in the $R$-parity conserving 
supersymmetry models.
KK parity conservation also implies
that the contributions to various low-energy
observables only arise at loop level and are small.
As a result, the limits on the scale $R^{-1}$ of the extra dimension
from precision electroweak data are rather weak, constraining  $R^{-1}$ to 
be larger than approximately 250~GeV. 
An attractive feature of UED models with KK parity 
is the presence of a stable massive particle which can be
a cold dark matter candidate 
\cite{Servant:2002aq,Burnell:2005hm,Kong:2005hn,Cheng:2002ej}.

\section{Kaluza-Klein Dark Matter}

The first and only comprehensive calculation of the UED relic density 
to date was performed in~\cite{Servant:2002aq}. 
The authors considered two cases of LKP: 
the KK hypercharge gauge boson $B_1$ and the KK neutrino $\nu_1$. 
The case of $B_1$ LKP is naturally obtained in Minimal UED (MUED)~\cite{Cheng:2002ab}, where the
radiative corrections to $B_1$ are the smallest in size,
since they are only due to hypercharge interactions.
The authors of \cite{Servant:2002aq} also realized the
importance of coannihilation processes and included
in their analysis coannihilations with the
$SU(2)_W$-singlet KK leptons, which in MUED 
are the lightest among the remaining $n=1$ KK particles.
It was therefore expected that their coannihilations will be most important.
Subsequently, Ref.~\cite{Kakizaki:2005uy} analyzed
the resonant enhancement of the $n=1$ (co)annihilation cross-sections
due to $n=2$ KK particles and ref.~\cite{Shah:2006gs} considered the influence 
of gravitons on the final relic density results.
Here we complete the LKP relic density calculation of Ref.~\cite{Servant:2002aq} and summarize our result.
%
%
\begin{figure}[t]
\includegraphics[width=5cm]{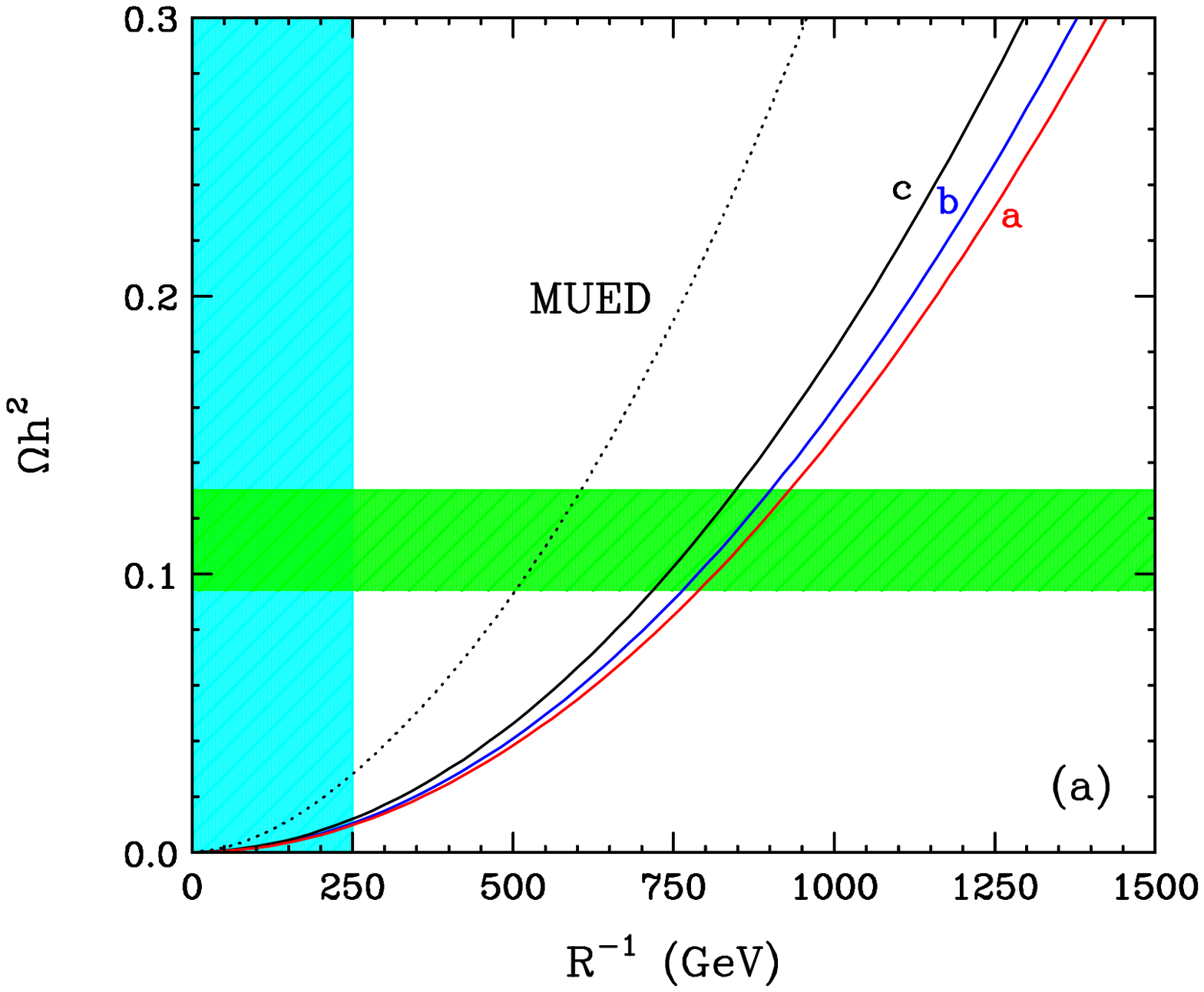}
\includegraphics[width=5cm]{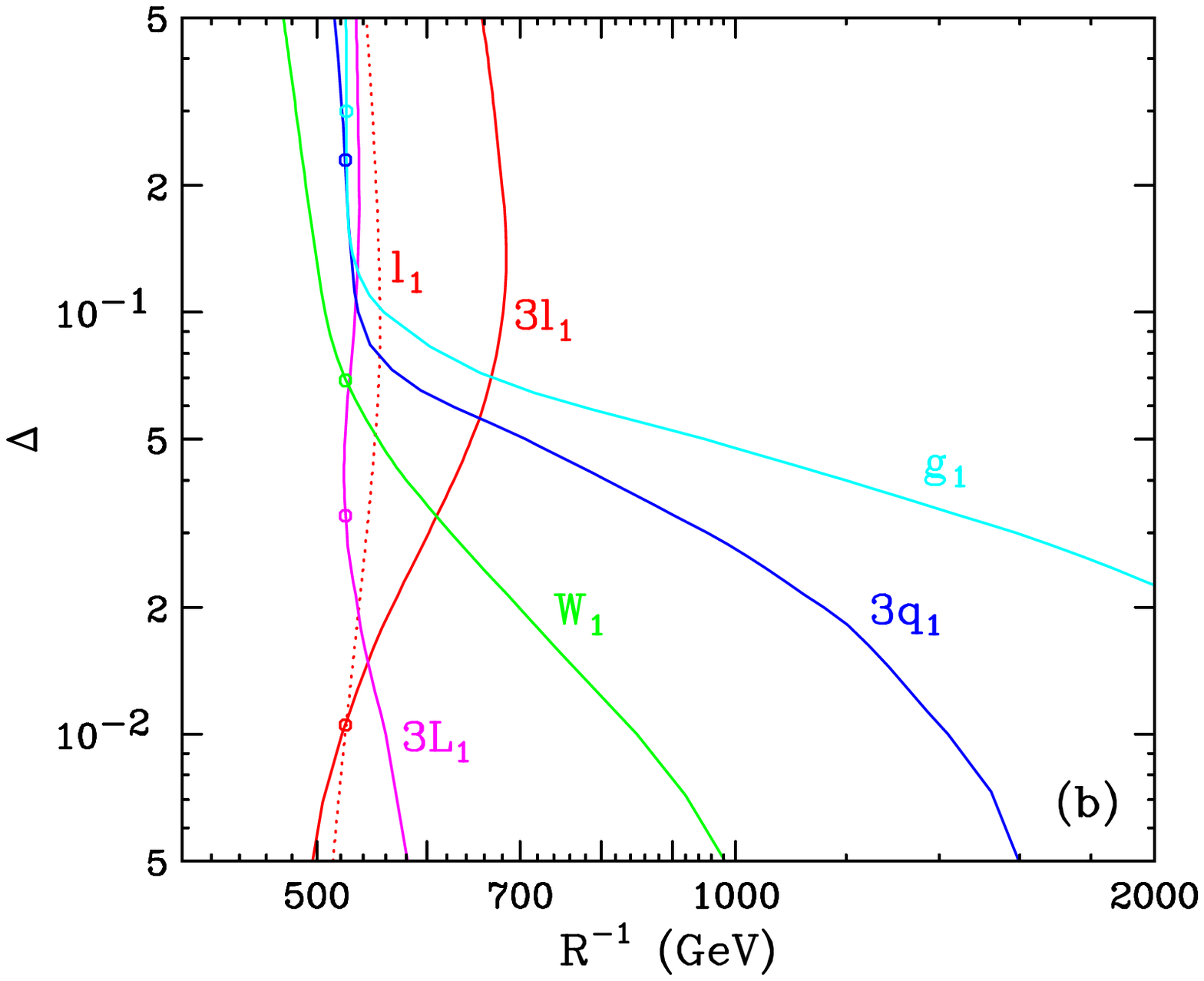}
\includegraphics[width=4.6cm]{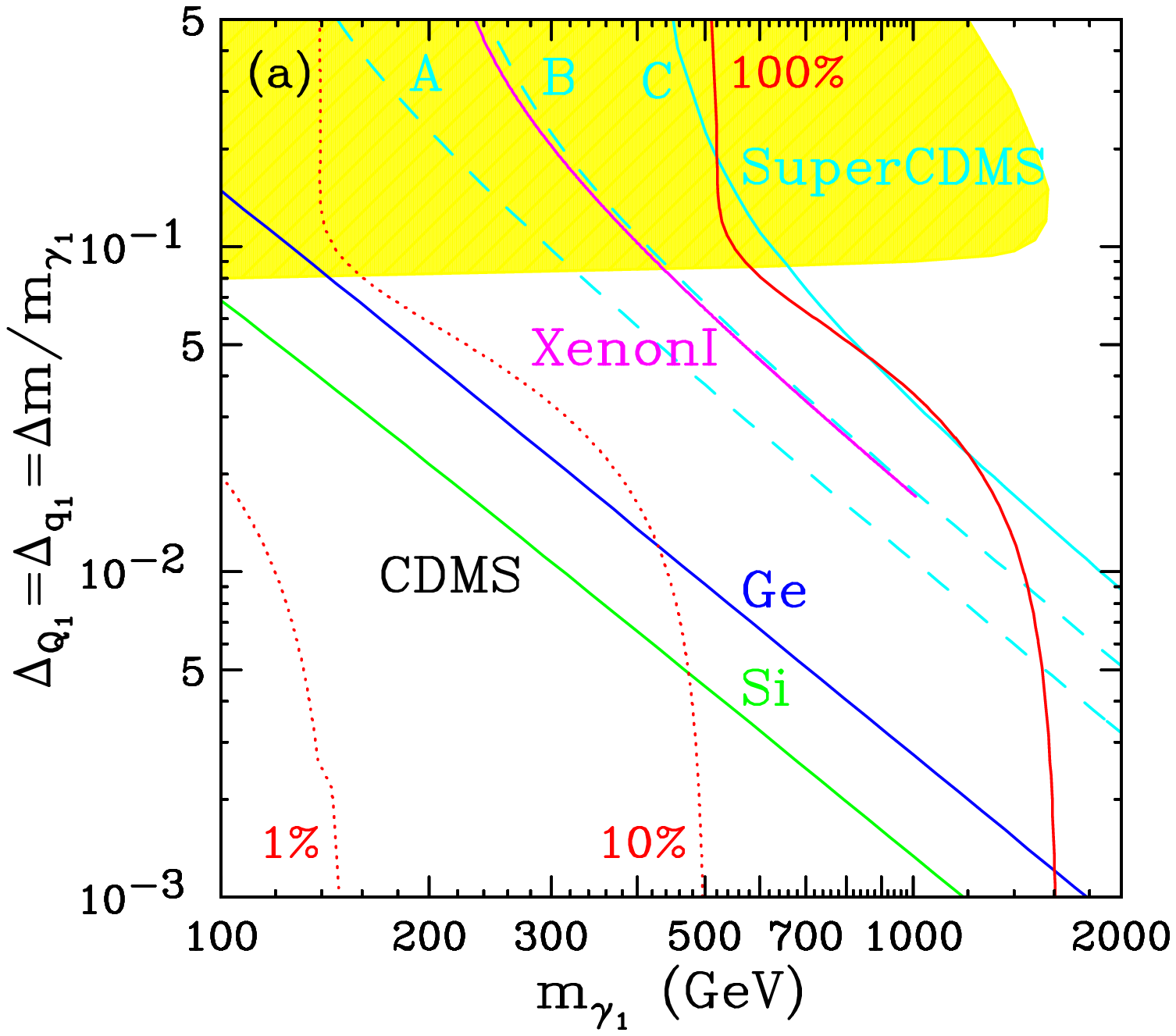}
\caption{\sl (a) Relic density of the LKP as a function of $R^{-1}$ in the MUED model. 
(b) The change in the cosmologically preferred value for $R^{-1}$
as a result of varying the different KK masses away from their nominal 
MUED values.
(c) The spin-independent direct detection limit from CDMS experiment for $\gamma_1$}.
\label{fig:KKDM}
\end{figure}
%
%
Fig.~\ref{fig:KKDM}(a) shows the relic density of the LKP as a function of $R^{-1}$ 
in the Minimal UED model. The (red) line marked ``a'' is the result
from considering $\gamma_1\gamma_1$ annihilation only, following
the analysis of Ref.~\cite{Servant:2002aq}, 
assuming a degenerate KK mass spectrum. The 
(blue) line marked ``b'' repeats the same analysis, but
uses $T$-dependent, effectively massless degrees of freedom
and includes the relativistic correction to the $b$-term in the non-relativistic velocity expansion. 
The (black) line marked ``c'' relaxes the assumption of 
KK mass degeneracy, and uses the actual MUED mass spectrum. The dotted line is
the result from the full calculation in MUED, including all coannihilation 
processes, with the proper choice of masses. The green horizontal band
denotes the preferred WMAP region for the relic density 
$0.094<\Omega_{CDM}h^2<0.129$. The cyan vertical band delineates
values of $R^{-1}$ disfavored by precision data.

Fig.~\ref{fig:KKDM}(b) shows the change in the cosmologically preferred value for $R^{-1}$
as a result of varying the different KK masses away from their nominal 
MUED values. Along each line, the LKP relic density is $\Omega_\chi h^2=0.1$.
To draw the lines, we first fix the MUED spectrum, and then vary the 
corresponding KK mass and plot the value of $R^{-1}$ which is required
to give $\Omega_\chi h^2=0.1$. We show variations of the masses of
one (red dotted) or three (red solid) generations of $SU(2)_W$-singlet
KK leptons; three generations of $SU(2)_W$-doublet leptons (magenta);
three generations of $SU(2)_W$-singlet quarks (blue)
(the result for three generations of $SU(2)_W$-doublet quarks
is almost identical); KK gluons (cyan) and
electroweak KK gauge bosons (green). The circle on each
line denotes the MUED values of $\Delta$ and $R^{-1}$.
%
%
%
%

The spin-independent direct detection limit from CDMS experiment is shown in fig.~\ref{fig:KKDM}(c).
We show the relic density and spin-independent direct detection limit from CDMS experiment 
in the plane of mass splitting 
$\Delta_{Q_1} = \Delta_{q_1} = \frac{m_{Q_1}-m_{\gamma_1}}{m_{\gamma_1}}$ and LKP mass for 
$\gamma_1$ LKP. The red line accounts for all of the dark matter (100\%) and 
the two red dotted lines show 10\% and 1\%, respectively. The blue (green) line shows 
the current CDMS limit with Ge-detector (Si-detector) and 
the three cyan lines represent projected SuperCDMS limits for each phase: A (25 kg), B (150 kg) and C (1 ton) 
respectively. In the case of $\gamma_1$ LKP, SuperCDMS rules out most of parameter space. 
The yellow region in the case of $\gamma_1$ LKP shows 
parameter space that could be covered by the collider search in $4\ell+\met$ channel at the LHC~\cite{Cheng:2002ab}.

\section{Discrimination of SUSY and UED}

We see that while $R$-parity conserving SUSY implies 
a missing energy signal, the reverse is not true: a missing energy
signal would appear in any model with a dark matter candidate, 
and even in models which have nothing to do with the dark 
matter issue, but simply contain new neutral quasi-stable particles.
Similarly, the equality of the couplings 
is a celebrated test of SUSY. It is only a necessary, 
but not a sufficient condition in proving supersymmetry. 
We are therefore forced to concentrate on discrimination between SUSY and UED.
There are two fundamental distinctions between them.
Let us begin with feature 1: the number of new particles.
The KK particles at $n=1$ are analogous to superpartners 
in supersymmetry~\cite{Cheng:2002ab}. The particles at the higher 
KK levels have no analogues in $N=1$ supersymmetric models.
Discovering the $n\ge2$ levels of the KK tower would therefore 
indicate the presence of extra dimensions rather than SUSY.
However these KK particles can be too heavy to be observed. 
Even if they can be observed at the LHC, they can be confused with 
other new particles~\cite{Battaglia:2005ma,Datta:2005zs} such as $Z'$ or 
different types of resonances from extra dimensions~\cite{Burdman:2006gy}.
The discovery opportunities for the $n=2$ level at the LHC and the Tevatron are discussed in~\cite{Datta:2005zs} 
(for linear collider studies of $n=2$ KK gauge bosons, see~\cite{Battaglia:2005ma,Battaglia:2005zf}).

\begin{figure}[t]
\includegraphics[width=5cm,height=4cm]{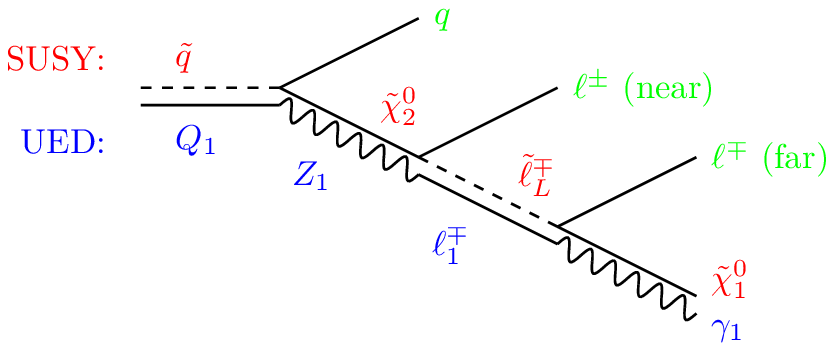}
\includegraphics[width=5cm]{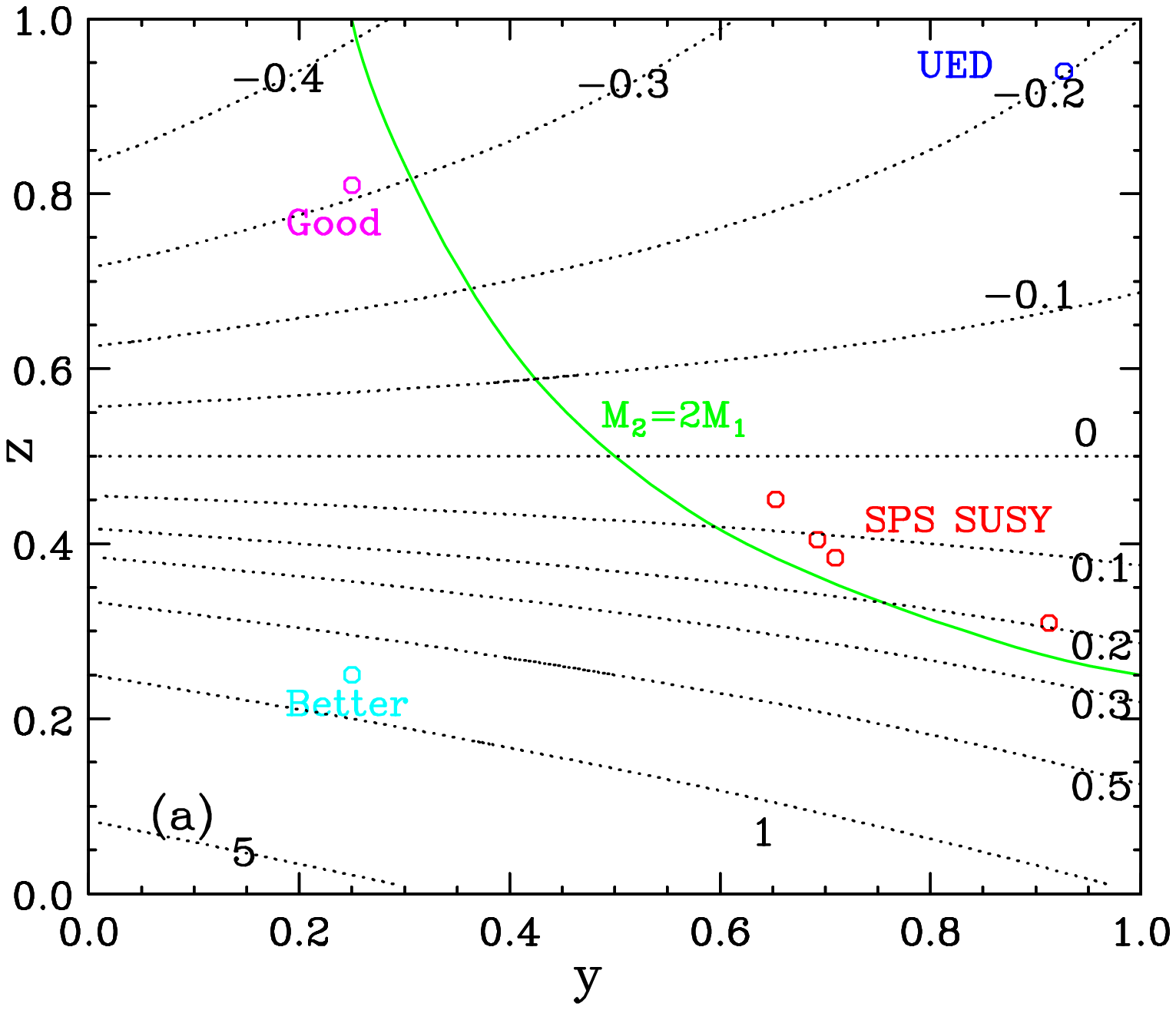}
\includegraphics[width=5.2cm]{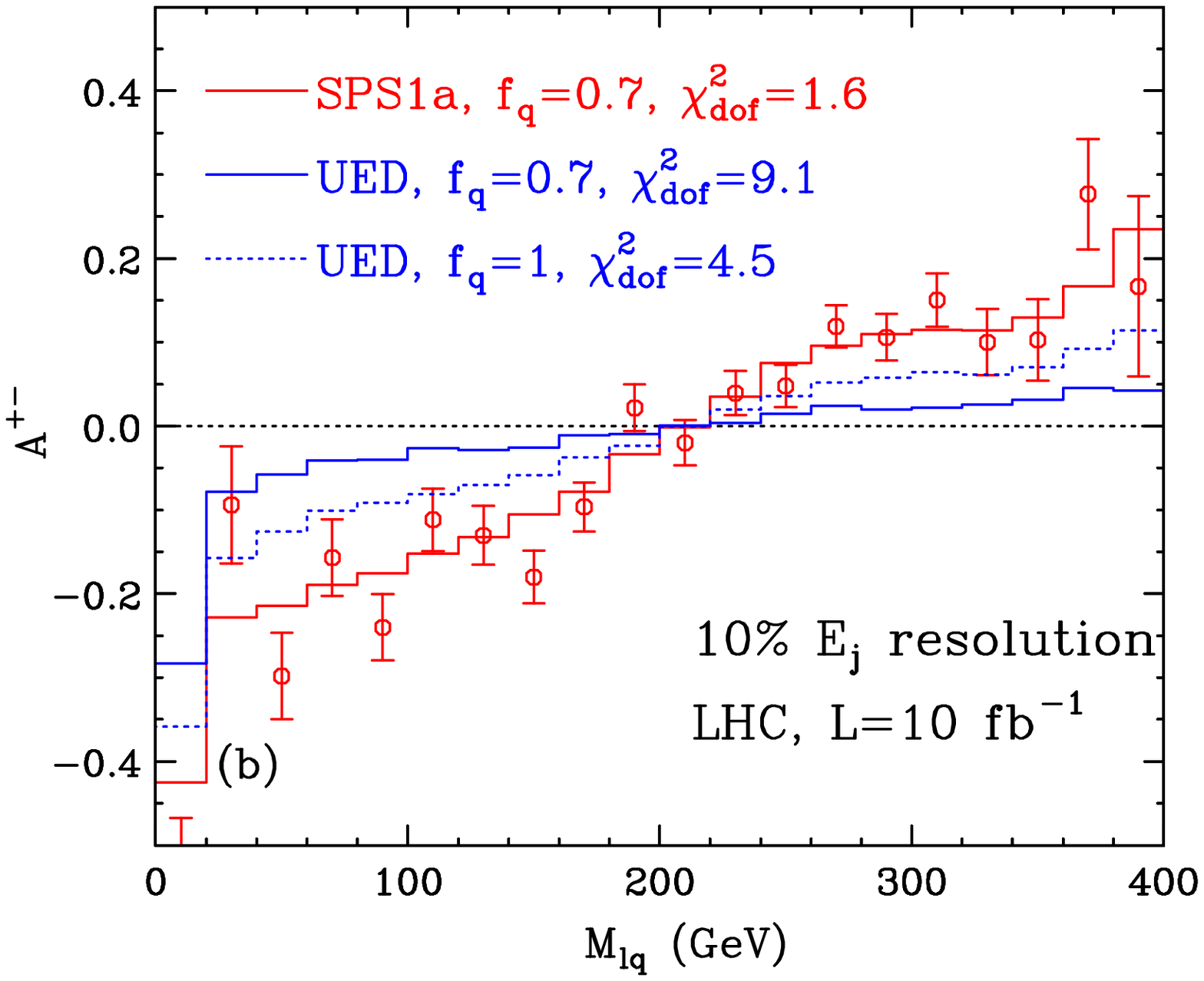}
\caption{\sl Spin determinations at the LHC using (a) the dilepton mass and (b) the asymmetry}
\label{fig:spins}
\end{figure}

The second feature -- the spins of the new particles -- 
also provides a tool for discrimination between SUSY and UED. 
Recently it has been suggested that a charge asymmetry in the lepton-jet invariant mass
distributions from a particular cascade, can be used to discriminate
SUSY from the case of pure phase space decays~\cite{Barr:2004ze}
and is an indirect indication of the superparticle spins.
It is therefore natural to ask whether this method can be
extended to the case of SUSY versus UED discrimination~\cite{Battaglia:2005ma,Datta:2005zs,Smillie:2005ar}. 
Following~\cite{Barr:2004ze}, we concentrate on
the cascade decay $\tilde q \to q\tilde\chi^0_2 \to q\ell^\pm\tilde\ell^\mp_L
\to q\ell^+\ell^-\tilde\chi^0_1$ in SUSY and the analogous decay chain
$Q_1 \to q Z_1\to q\ell^\pm\ell^\mp_1\to q\ell^+\ell^-\gamma_1$ in UED (see fig.~\ref{fig:spins}(a)). 
%
%
The invariant mass distributions for SUSY/Phase space can be written as $\frac{dN}{d  \hat{m} } = 2 \hat{m}$, 
while for UED it is $\frac{dN}{d  \hat{m} } = \frac{4(y+4z)}{(1+2z)(2+y)} \left ( \hat{m} + r \, \hat{m}^3\right )$
~\cite{Smillie:2005ar,KK}.
The coefficient $r$ in the second term of the UED distribution is defined as $r = \frac{(2-y)(1-2z)}{y+4z}$, 
$\hat{m} = \frac{m_{\ell\ell}}{m_{\ell\ell}^{max}}$ is the rescaled invariant mass, 
$y = \left ( \frac{m_{\tilde\ell}}{m_{\tilde{\chi}_2^0}} \right )^2$ and 
$z = \left ( \frac{m_{\tilde{\chi}_1^0}}{m_{\tilde\ell}} \right )^2$ are 
the ratios of the masses involved in the decay. $y$ and $z$ are less than 1 in the case of on-shell decay.
We see that whether or not the UED distribution is the same as the SUSY distribution depends on 
the size of the coefficient $r$ in the second term of the UED distribution. 
The UED distribution becomes exactly the same as the SUSY distribution if $r=0.5$.
Therefore we scan the $(y, z)$ parameter space, calculate the coefficient $r$ and 
show our result in fig.~\ref{fig:spins}(a). 
In fig~\ref{fig:spins}(a) contour dotted lines represent the size of the coefficient $r$.
The minimal UED case is denoted by the blue dot in the upper-right corner 
since $y$ and $z$ are almost 1 due to the mass degeneracy. 
The red dots represent several snowmass points: SPS1a, SPS1b, SPS5 and SPS3 from left to right.
The green line represents gaugino unification 
so all SUSY benchmark points are close to this green line. 
In fig.~\ref{fig:spins}(b), we generated data samples from SPS1a assuming $10fb^{-1}$ and 
constructed the asymmetries in SUSY and UED.
We included 10\% jet energy resolution.
Red dots represent data points, the red line is the SUSY fit to the data points and 
the blue lines are the UED fits to the data points for two different $f_q$'s. 
$\chi^2$-minimized UED (SUSY) 
fits to data are shown in blue (red).
For SUSY, $\chi^2$ is around 1 as we expect. 
We can get better $\chi^2$ for UED from 9.1 to 4.5 by increasing $f_q$. 
It is still too big to fit the experimental data. 
So our conclusion for this study is that a particular point like SPS1a can not be faked through
the entire parameter space of UED.
However we need to check whether this conclusion will remain the same when we include the wrong jets 
which have nothing to do with this decay chain.
\begin{theacknowledgments}
The work of KK and KM is supported in part by 
a US Department of Energy Outstanding Junior Investigator 
award under grant DE-FG02-97ER41209.
\end{theacknowledgments}

\end{document}